\documentclass[lettersize,journal]{IEEEtran}
\usepackage{amsmath,amsfonts}
\usepackage{algorithmic}
\usepackage{algorithm}
\usepackage{array}
\usepackage[caption=false,font=normalsize,labelfont=sf,textfont=sf]{subfig}
\usepackage{textcomp}
\usepackage{stfloats}
\usepackage{xurl}
\usepackage{verbatim}
\usepackage{graphicx}
\usepackage{cite}
\usepackage{multicol}
\usepackage{xcolor}
\usepackage{colortbl}
\usepackage{hyperref}
\hypersetup{colorlinks, citecolor=green, filecolor=pink, linkcolor=blue, urlcolor=blue }

\begin{document}

\title{Towards Green Metaverse Networking: Technologies, Advancements and Future Directions}
\author{Siyue Zhang, Wei Yang Bryan Lim, Wei Chong Ng, Zehui Xiong, Dusit Niyato,~\IEEEmembership{IEEE Fellow}, \\ Xuemin Sherman Shen,~\IEEEmembership{IEEE Fellow}, Chunyan Miao,~\IEEEmembership{IEEE Fellow}}

\maketitle

\begin{abstract}

As the Metaverse is iteratively being defined, its potential to unleash the next wave of digital disruption and create real-life value becomes increasingly clear. With distinctive features of immersive experience, simultaneous interactivity, and user agency, the Metaverse has the capability to transform all walks of life. However, the enabling technologies of the Metaverse, i.e., digital twin, artificial intelligence, blockchain, and extended reality, are known to be energy-hungry, therefore raising concerns about the sustainability of its large-scale deployment and development. This article proposes Green Metaverse Networking for the first time to optimize energy efficiencies of all network components for Metaverse sustainable development. We first analyze energy consumption, efficiency, and sustainability of energy-intensive technologies in the Metaverse. Next, focusing on computation and networking, we present major advancements related to energy efficiency and their integration into the Metaverse. A case study of energy conservation by incorporating semantic communication and stochastic resource allocation in the Metaverse is presented. Finally, we outline the critical challenges of Metaverse sustainable development, thereby indicating potential directions of future research towards the green Metaverse. 

\end{abstract}


\section{Introduction}

The Metaverse is arriving at an inflection point of development thirty years after the term was first coined in Neal Stephenson's science-fiction novel "Snow Crash". While its definition is still fluid today, the Metaverse is generally regarded as an embodied version of the Internet that seamlessly synergizes virtual and physical worlds, and enables users to live and interact in virtual worlds through controlling personalized avatars.

A growing amount of commitment, investment, and effort has been made to realize the Metaverse across academia and industry. For the former, the number of research publications related to the Metaverse in 2022 is five times more than that in 2021. For the latter, Facebook was rebranded as Meta to shift from a media company to a Metaverse  company in 2021. Such excitement about the Metaverse is driven by improved technical and marketplace readiness. For example, the enabling technologies of the Metaverse have undergone rapid advancements recently, including the Internet-of-Things (IoT), digital twin (DT), 5G/6G, artificial intelligence (AI), blockchain, virtual reality (VR), and augmented reality (AR). Moreover, the COVID-19 pandemic has accustomed users to embracing digital life, e.g., online meetings and shopping.

As compared to the massively multiplayer online role-playing games (MMORPGs) of today, the Metaverse is a far broader concept that features user-generated content (UGC), a variety of use cases, decentralization, and interoperability across platforms and devices. However, there are still technical deficiencies in communication speed, computing power, data storage, etc., to fully realize the envisioned Metaverse, not to mention social barriers. As the Metaverse is relatively nascent, most research has been centered on Metaverse architecture, enabling technologies, and applications. The study of \cite{edgeMetaverse}  discusses the Metaverse implementation at mobile edge networks at scale. It presents a visionary four-layer architecture positioning enabling technologies in the edge-enabled Metaverse. In addition to technology, \cite{allone} addresses social aspects of the Metaverse, e.g., acceptance, security, and privacy. 

While the sustainability aspect has been pointed out in the aforementioned studies, no study has yet to evaluate the energy consumption, efficiency, and sustainability of the Metaverse. The energy demand could easily escalate as the Metaverse will require 1000-fold more computing power and push data usage by 20 times in the next 10 years. \footnote{\url{https://www.intel.com/content/www/us/en/newsroom/opinion/powering-metaverse.html}}\textsuperscript{,}\footnote{\url{https://techblog.comsoc.org/2022/02/20/credit-suisse-metaverse-to-push-data-usage-by-20-times-worldwide-by-2032/}} From our perspective, Metaverse sustainability is the ability of the Metaverse to exist and develop without depleting natural resources for the future, e.g., fossil fuels for energy and raw materials for electronic devices. 
It is becoming particularly critical due to increasing economic and environmental costs, user awareness, and stringent sustainability targets such as the Net Zero coalition.\footnote{\url{https://www.un.org/en/climatechange/net-zero-coalition}} Therefore, this article makes the first effort to optimize energy efficiencies of all network components for realizing the Metaverse, i.e., Green Metaverse Networking (GMN). The article provides an analytical framework in a technology-based approach for evaluating energy efficiency. Specifically, our framework centers around the four main groups of energy-hungry enabling technologies namely:
\begin{itemize}
    \item \textit{Digital Twin} about creating and operating the Metaverse,
    \item \textit{Artificial Intelligence} about endowing the Metaverse with autonomy and intelligence,
    \item \textit{Extended Reality} about user interaction with the Metaverse, and
    \item \textit{Blockchain} about data storage and interoperability in the Metaverse.
\end{itemize}

The structure and key novelties of this article are as follows:
\begin{enumerate}
    \item We identify energy-hungry technologies and their application scenarios, analyze their consumption, efficiency, and influencing factors, and outline sustainability-related metrics, thereby providing an analytical framework for assessing the Metaverse sustainability in Section \ref{sec2}.
    \item We review the forefront computing and networking advancements of these technologies towards GMN in Section \ref{sec3}. Through a case study in Section \ref{sec4}, we illustrate how semantic communication and stochastic resource allocation can contribute to energy conservation, thereby guiding the design of future communication systems for the Metaverse.
    \item We outline the critical challenges and future research directions towards GMN in Section \ref{sec5}.
\end{enumerate}

\begin{figure*}[hb]
\centering
\includegraphics[width=7in]{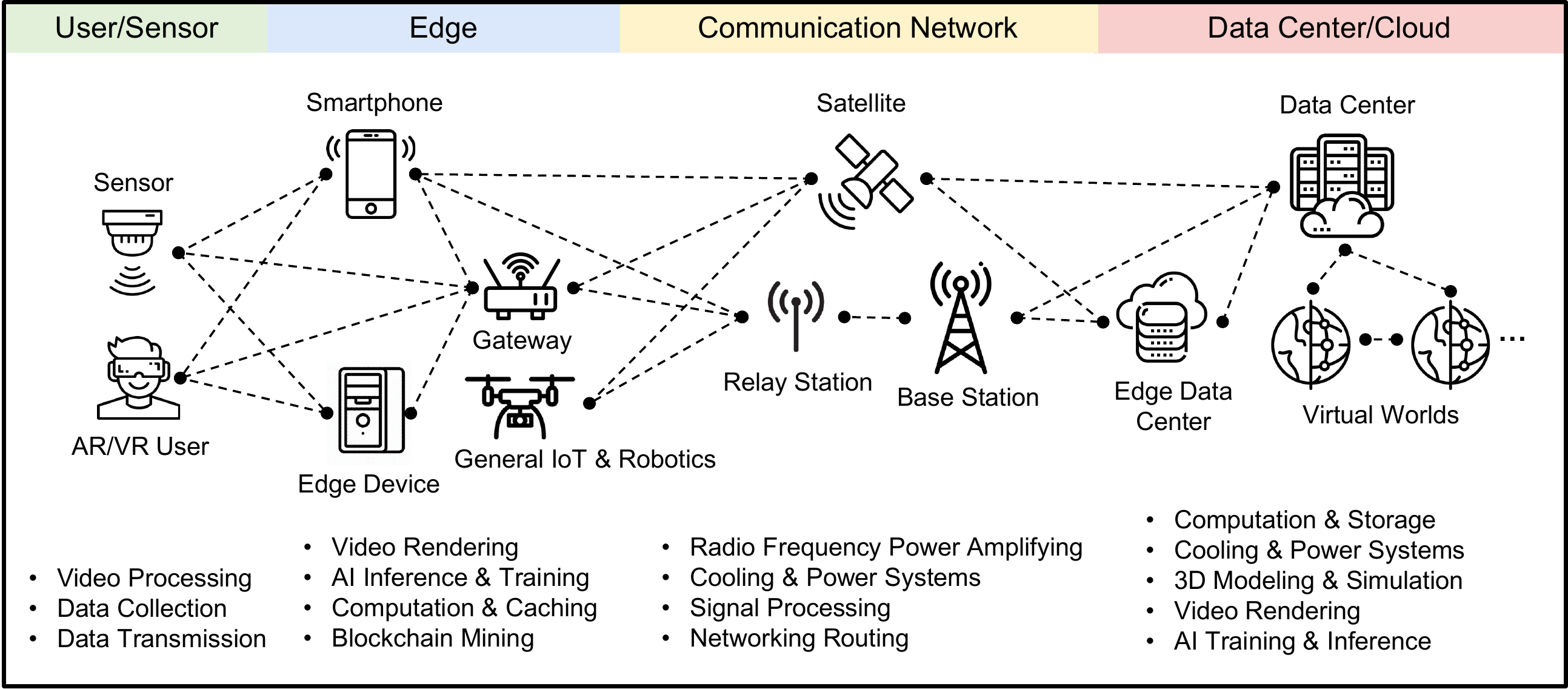}
\caption{A four-tier Metaverse architecture and energy-intensive processes.}
\label{fig_arc}
\end{figure*}

\section{Energy-Hungry Technologies, Energy Consumption, Efficiency and Metrics}
\label{sec2}

    Metaverse is made possible due to the development of enabling technologies. But many of them are energy-hungry. Their sustainability-related metrics are summarized in Table \ref{tab:table1}.

    \subsection{Digital Twin}

    The DT is a digital representation that accurately reflects the physical world. It provides real-time visibility, data analytics, and the capability to predict potential changes in the physical world. As the backbone of DT, IoT collects physical data by networked sensors, derives insights from data by edge processors, transmits insights by gateways, and proceeds Metaverse commands by executors (Fig. \ref{fig_arc}). While IoT has already connected physical objects to the Internet nowadays, the Metaverse will spur the demand for more advanced and interconnected IoT devices (e.g., AR/VR head-mounted displays (HMDs) and unmanned aerial vehicles (UAVs)) to enable seamless synchronization between physical and digital worlds. Undoubtedly, the increasing quantity and complexity of IoT devices would impose challenges on sustainable energy supply, especially since most devices are wireless and powered by limited batteries.
    
    Network efficiency has a considerable influence on DT energy consumption, including radio networks, IoT, wireless sensor networks (WSNs), etc. The efficiency varies depending on the network device, communication mode, protocol, and software implementation. Typically, network efficiency is measured by the average power consumption at a certain level of service quality. The Quality of Service (QoS) metrics such as throughput, latency, and packet loss are well-defined to evaluate classical information-theoretic network performance. However, the perceived quality by users could be of more importance, especially for content providers. For example for video streaming, users could be sensitive to quality degradation due to varying bit rates although the throughput is high. Maintaining the bit rate at a lower but stable level, i.e. lower QoS, could improve user experience and reduce energy consumption at the same time. Therefore, a paradigm shift towards quality of experience (QoE) is called by content-centric Metaverse applications to focus on the overall acceptability of service perceived by the user.
    
    For wireless communications, the energy efficiency is addressed by data rate per power (in bit/J), which is currently at the order of 10 Mbit/J. Spectral efficiency (in bps/Hz) measures the data throughput over a given spectrum for a site. Area energy efficiency (in bps/Hz/m$^2$/J) is the primary energy efficiency metric for current wireless systems, which equals spectral efficiency divided by area coverage and energy consumption. 5G pledges 10 times higher area energy efficiency than 4G. Along with the rapid emergence of aerial nodes such as UAVs, volumetric energy efficiency (in bps/Hz/m$^3$/J) would become a more important metric in future systems. 
    
    Apart from networking, a significant amount of energy is required for computation. For example, 3D modeling and simulation are essential for creating avatars, DTs, and virtual worlds. To create immersive and realistic experiences, 3D models are developed with computation-intensive simulations of physics, lighting, sound, and other elements. Compared to creating virtual worlds, around-the-clock server hosting and operations in data centers consume far more energy because of numerous user connections, heavy data traffic, and intensive computational load. Cloud-native solutions are required to offload compute-intensive tasks (e.g., 3D rendering and AI training) to high-end servers in remote data centers when the capability of the local computer is limited. But the over-dependence on cloud computing could aggravate energy consumption. Taking cloud gaming as an example, the power demand in the upstream network together with data center graphic processing was estimated to be markedly (e.g., 40-60\% for desktops and 120-300\% for laptops) higher than that of local gaming \cite{game}. 

    Inevitably, the predominant cloud-based solutions of today lead to exacerbated energy challenges. Three major components contribute to the energy usage of data centers: physical servers, cooling systems, and network devices. They usually consume 40–55\%, 15–30\%, and 10–25\% of the total power respectively. To offset the strong growth in energy demand, remarkable efficiency improvements were implemented in servers, storage, network switches, and infrastructure design. The most efficient data centers (e.g., Google) nowadays reach a power usage effectiveness (PUE) of $\sim$1.1, while the global average of large data centers is $\sim$1.57. \footnote{\url{https://www.google.com/about/datacenters/efficiency/}}

    \subsection{Artificial Intelligence}
        
    AI plays an important role in endowing the Metaverse with autonomy and intelligence, thereby improving the QoE of users. For example, Computer Vision (CV), which derives meaningful information from images and videos, enables the Metaverse to provide services such as object detection, activity recognition, avatar navigation, and visual reasoning. Natural Language Processing (NLP) enables conversational non-playable characters (NPCs) in the Metaverse and also provides services including language translation, text summarization, sentiment analysis, and speech recognition. Generative AI utilizes existing text, audio files, or images to create 3D objects, digital assets (e.g., music, painting, article), user avatars, NPCs, and even whole virtual worlds. Beyond creating innovative services, AI can be integrated with other technologies to provide more intelligent services, e.g., fast 3D rendering, efficient communication, and intelligent blockchains. 
    
    Despite widespread applications, AI comes at a high cost economically and environmentally due to high energy consumption. In AI training, computers are iteratively fed with training data and calculate the ML parameters continuously. Around 1000 MWh of electricity is spent to train a single language model like OpenAI GPT-3. An exponential increase of compute used in the largest AI training was observed in the last decade with a 3.4-month doubling time.\footnote{\url{https://openai.com/blog/ai-and-compute/}} Fortunately, unlike cloud-centric AI training, AI inference is trending to be shifted to distributed edge devices for higher scalability, lower operational risks, and reduced latency. However, this results in the additional challenges of limited processing power and energy supply of edge devices.

    To evaluate AI technologies, both accuracy and energy should be considered. Accuracy metrics are well-defined and easily computed for specific tasks. However, energy is rarely measurable and difficult to compute. Hence, a few alternatives have been used. The model size measures the ML model's memory consumption, which is strongly dependent on the algorithm and model architecture. The total number of floating-point operations (FLOPs) is used to approximate the number of multiply-and-accumulate (MAC) operations, indicate the computational workload, and imply energy consumption. Nonetheless, both metrics do not consider the impact of data movement between the memory unit and the processing unit, which may dominate the energy consumption in the computation \cite{sze2020evaluate}. Thus, a few energy estimation methodologies were developed for precise energy estimation for specific ML models and hardware devices. The trade-offs between accuracy and energy are necessary for AI training considering the diminishing marginal return of accuracy when increasing the dataset size and model complexity. It is especially true for distributed AI training, e.g., federated learning (FL), when increasing the concurrency.
 
\begin{table*}[!t]
\caption{Summary of Metaverse Sustainability-Related Metrics\label{tab:table1}}
\renewcommand{\arraystretch}{1.3} % Default value: 1
\setlength{\tabcolsep}{5pt} % Default value: 6pt
\centering
\begin{tabular}{|l|l|l|l|}
\hline
\textbf{Digital Twin} & \textbf{Consumption Point}  & \textbf{Unit} & \textbf{Description}\\
\hline
\hline
Power utilization effectiveness (PUE) & Data Center  & $\mathbb{R^+}\in(1,\infty)$ & ratio of total facility energy to IT equipment energy\\
\hline
Data center energy productivity (DCeP) & Data Center & $\mathbb{R^+}\in(0,1)$ & ratio of useful work produced to total energy\\
\hline
IT equipment utilization (ITEU) & Data Center  & $\mathbb{R^+}\in(0,1)$ & ratio of actual IT power to total rated IT power\\
\hline
Carbon usage effectiveness (CUE) & Data Center & kgCO$_2$/kWh & ratio of CO$_2$ emissions to IT equipment energy\\
\hline
Green energy coefficient (GEC) & Data Center & $\mathbb{R^+}\in(0,1)$ & ratio of green energy to total energy consumption\\
\hline
Power at QoS/QoE/QoPE & Wireless system  & W  & average power at a certain level of quality\\
\hline
Energy efficiency (EE) & Wireless system  & bit/J  & ratio of data rate to required power\\
\hline
Energy consumption rating (ECR) & Wireless system  & W/Gbps & ratio of aggregated energy consumption to capacity\\
\hline
Area Power Consumption (APC) & Wireless system & W/m$^2$  & ratio of power consumption to served area \\
\hline
Spectral efficiency & Wireless system  & bps/Hz  & ratio of information transmission rate to bandwidth\\
\hline
Area energy efficiency & Wireless system  & bps/Hz/m$^2$/J  & ratio of area spectral efficiency to energy consumption\\
\hline
Volumetric energy efficiency & Wireless system  & bps/Hz/m$^3$/J  & ratio of space spectral efficiency to energy consumption\\
\hline
Network level performance indicator & Wireless system  & subscribers/W  & ratio of number of subscribers on busy hours to power\\
\hline
\arrayrulecolor{white}\hline
\\
\arrayrulecolor{black}\hline
\textbf{Artificial Intelligence} & \textbf{Consumption Point} & \textbf{Unit} & \textbf{Description}\\
\hline
\hline
Model size & ML model & bytes & storage space taken by ML model\\
\hline
Number of parameters & ML model & $\mathbb{N^+}\in(1,\infty)$ & number of learnable parameters in ML model\\
\hline
Elapsed time & ML model  & s & running time for generating an AI result\\
\hline
FLOPs & ML model  & FLOP & number of floating-point operations\\
\hline
FLOPS per watt & Processor & FLOPS/W & ratio of number of floating-point operations to energy\\
\hline
\arrayrulecolor{white}\hline
\\
\arrayrulecolor{black}\hline
\textbf{Blockchain} & \textbf{Consumption Point} & \textbf{Unit} & \textbf{Description}\\
\hline
\hline
Overall annual energy & Blockchain  & TWh/yr & overall yearly energy consumption of the network\\
\hline
Energy per transaction & Blockchain & kWh/txn & ratio of overall energy to the number of transactions\\
\hline
Coin carbon footprint & Blockchain  & kgCO$_2$/coin & ratio of carbon emissions to number of coins mined\\
\hline
Transaction carbon footprint & Blockchain  & kgCO$_2$/txn & ratio of carbon emissions to number of transactions\\
\hline
\arrayrulecolor{white}\hline
\\
\arrayrulecolor{black}\hline
\textbf{Extended Reality} & \textbf{Consumption Point} & \textbf{Unit} & \textbf{Description}\\\hline
\hline
Power at QoS/QoE/QoPE & HMD  & W & average power at a certain level of quality\\
\hline

\end{tabular}
\end{table*}

    \subsection{Extended Reality}

    XR extends physical reality into virtual reality to different extents: AR overlays virtual objects on the physical world and VR simulates a fully virtual world. The emergence of XR revolutionizes the Metaverse by creating brand-new virtual experiences in any imaginable domain. To have an immersive sensory experience, users wear a standalone HMD (e.g., Oculus Quest) or an HMD connected to a computer (e.g., Oculus Rift). High-end VR systems include sensors like accelerometers, gyroscopes, and magnetometers, which can collect more  physical world data. 360$^\circ$ 2D videos rendered from 3D models are transmitted to the edge device or HMD for local processing before displaying behind the eyes.   
    
    VR video rendering is fulfilled mainly in data centers and edge computers due to the stringent requirements of processing hardware. It is reported to consume twice as much power as conventional planar video rendering in \cite{vr360}. In HMDs, video processing expends the majority of the power budget. Memory access and computation are two main contributors to video processing energy consumption. Moreover, VR systems are increasingly used to execute additional functions, e.g., sensing data collection and processing, AI training and inference, and caching, further increasing the energy demand.

    VR QoS metrics mainly include latency, resolution, field of view (FOV), and frame per second (FPS). The power at QoS is typically used to compare energy efficiency. Nevertheless, a device with better QoS does not necessarily provide better visual quality. For example, perceptually high-quality VR videos do not require uniformly high-quality pixels (i.e., full resolution) with a wide FOV. QoE-based energy efficiency metrics encourage the development of VR systems with better visual quality and less power required. Detailed examples are illustrated in Section \ref{adv-xr}.
    
    \subsection{Blockchain}
    \label{blockchain-now}

    As a distributed database, the blockchain can be used to store the massive amount of data generated in the Metaverse, e.g., UGC and IoT data. It decentralizes storage, improves scalability, reduces data transmission, and ensures data security and privacy. The development of cross-chain protocols allows data exchange among blockchains and improves Metaverse interoperability. Non-Fungible Tokens (NFTs) are unique cryptographic tokens on blockchain indicating the ownership of digital or real-world items as well as transaction history. NFTs can ensure uniqueness, track ownership, automate trading through smart contracts, and establish trust among parties, which highly motivates the creation of UGC. In contrast, cryptocurrency is adopted as a fungible digital currency with economic value. It eliminates the need for central authorities, reduces transaction costs, enhances security and privacy, and simplifies fund transfers. 

    Nevertheless, most of the existing blockchain applications are suffering deficiencies in operation efficiency and  energy consumption. Annually, Bitcoin is estimated to consume 60-125 TWh of electrical energy, which is comparable to the yearly consumption of countries like Austria and Norway. A single transaction of Bitcoin requires enough electricity to supply an average size household in Germany for weeks, which is multiple orders of magnitude higher than traditional systems. Two factors largely determine blockchain energy consumption.

    \textbf{Consensus Mechanism}: Although there is a growing variety of consensus mechanisms, the majority of contemporary blockchain applications are based on Proof-of-Work (PoW) like Bitcoin. PoW involves network nodes solving computationally intensive puzzles and competing to create a new block, namely mining. Because miners with more powerful hardware have a higher probability of success, it creates an arms race among miners and generates electronic waste.
    
\begin{figure*}[b]
\centering
\includegraphics[width=7in]{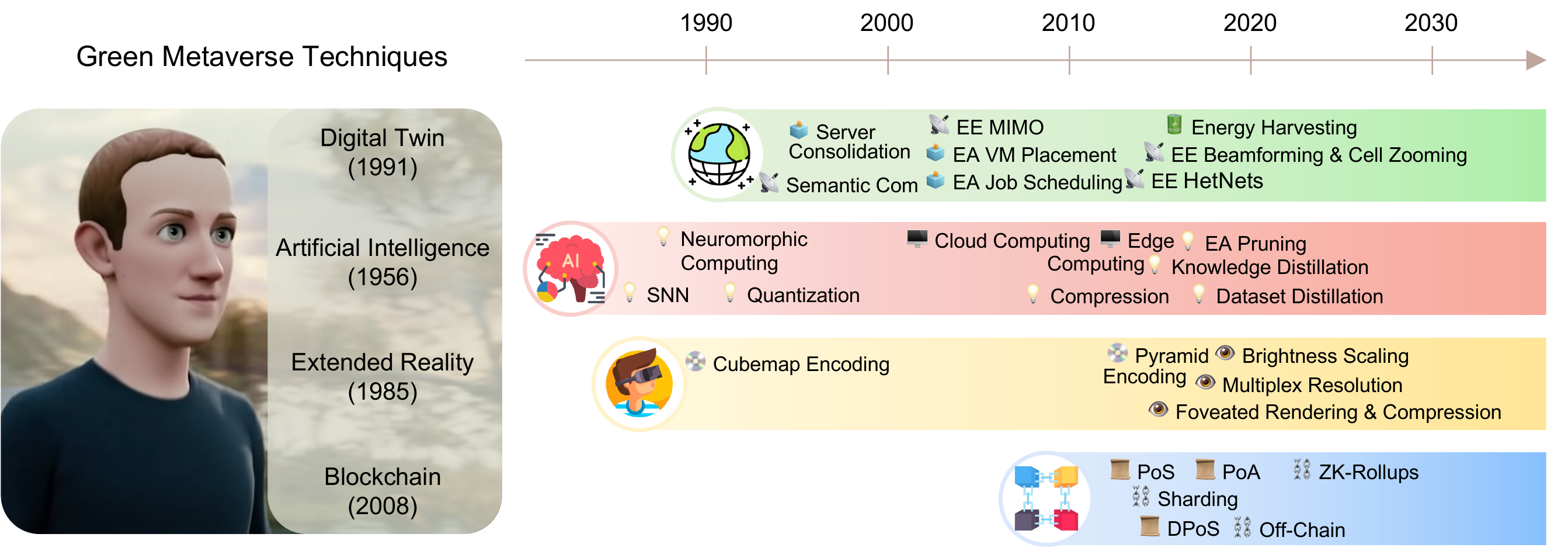}
\caption{Emerging and evolving Green Metaverse Techniques in history. (EE: energy-efficient, EA: energy-aware, Com: communication)}
\label{fig_techs}
\end{figure*}

    \textbf{Operation Redundancy}: Regardless of the type of consensus mechanisms, the block is continuously replicated across blockchain network nodes. The required computation and thus power consumption for data operations could be problematic when the number of nodes in the blockchain network rises rapidly resulting in a high degree of redundancy. Compared to centralized databases with a designed level of redundancy, a blockchain generally consumes much more than a corresponding centralized solution (by a factor of around 50) \cite{sedlmeir2021recent}.
    
    In a holistic view, the idle power consumption of nodes should also be considered apart from the above factors. To evaluate overall energy efficiency, energy consumption per transaction is commonly used to compare blockchains to other technologies. Based on this metric, most contemporary blockchain applications (e.g., Bitcoin) still spend a mammoth amount of energy compared to traditional payment systems (e.g., credit cards). Notably, this metric can be misleading because blockchain energy consumption largely depends on consensus mechanisms (e.g., mining in PoW) rather than transaction volume. In a nutshell, the overall energy consumption is still significant for current blockchains. The situation is expected to change after the world’s second-largest cryptocurrency, Ethereum, shifts from PoW to a greener proof-of-stake (PoS) system in Q4 2022.
    
    \textbf{Lessons learned:} Employing energy-inefficient technologies is detrimental to the Metaverse scalability and results in profound negative impacts on the environment. The metrics in Table \ref{tab:table1} are of great use to identify inefficient components in Metaverse networks. In many cases, trade-offs must be well-balanced between higher energy consumption and improved user experience. As the Metaverse will be human-centric, QoE-based energy-efficiency measures must be designed to directly link user experience with energy usage.

\section{Recent Technological Advancements in Energy Efficiency}
\label{sec3}

    Numerous research efforts have been put into individual technology for improving performance, potentially contributing to the Metaverse sustainability. The most significant improvements and their applications in the GMN are selected for discussion. A summary is presented in Fig. \ref{fig_techs}. 

    \subsection{Digital Twin}
    \label{adv-dt}

    \subsubsection{Sensor, User and Edge}

    In IoT systems, data collected by IoT devices are centralized either in an IoT gateway connected to the cloud or in a local IoT hub. The primary type of energy reduction measures focuses on IoT sensors. Because idle sensors consume unnecessary energy, an efficient duty scheduling algorithm is required to minimize power consumption while satisfying QoS requirements. The duty cycle can be designed beforehand based on predictions of communication periods. It can also be dynamically adjusted based on the real-time priority of traffic, device energy, and required power. 30-50\% energy reduction is obtained by optimizing the scheduling and workload distribution in \cite{efficientprotocols}. Moreover, novel sensing concepts including selective sensing, approximate sensing, and compressed sensing offer additional efficiency improvement in data transmission.
     
    IoT sensors can be powered by ambient free energy such as light, radio frequency, temperature difference, and kinetic energy, which is called energy harvesting. The research community has also worked towards developing energy-efficient routing protocols, especially energy harvest-based protocols for WSNs to cope with intermittent renewable resources. The energy-harvesting-aware routing algorithm proposed by \cite{harvestalgo} consumes 40-80\% less power and reduces 40\% packet loss compared to other algorithms by selecting the node with the highest residual energy level and the minimum cost link. Alternatively, edge devices can harvest energy provided by a wireless power transfer (WPT) system when ambient energy is not available. A novel algorithm developed by \cite{wu2022non} jointly optimizes the performance and energy consumption of the WPT-powered FL system with a significantly reduced computation time.
    
    \subsubsection{Communication Network}

    Unique features of accessibility, mobility, and scalability make wireless communication indispensable in the Metaverse.  A comprehensive review of future green radio networks has been conducted by \cite{green5g}.
    
    Heterogeneous networks (HetNets) engage diverse networks in a hierarchical structure to provide the service to the same user. Macrocells with a higher capacity provide services for the larger area coverage while small cells with better efficiencies are used to reduce energy consumption and enhance the network throughput. Optimizing cell selection and deployment promotes the HetNet energy efficiency gain. Cell zooming dynamically varies the cell size based on the traffic load, state of the channel, etc. When the cell is underutilized, it can transfer the traffic to nearby cells through cell zooming and enter sleep mode for energy conservation. Beamforming employs an array of antennas to focus electromagnetic radiation (EMR) toward the user's direction. So, the signal strength in a specific direction maximizes, EMR exposure and transmit power reduces \cite{green5g}. Massive multiple-input and multiple-output (mMIMO) is developed based on MIMO by packing more antennas into a small area. Compared to MIMO, mMIMO improves the capacity by 5 times and energy efficiency by 100 times due to the focus of radiated energy on user equipment \cite{green5g}. Device-to-Device (D2D) communication is a promising technique that enables direct communication between devices without traversing the network. Lastly, semantic communication is envisioned as a key component of 6G for reducing the transmission volume, facilitated by AI technologies. The extracted meanings of the information are transmitted rather than information symbols. A semantic communication case study is presented in Section \ref{sec4}.
    
    \subsubsection{Data Center and Cloud}

    An idle or underutilized server consumes two-thirds of the energy when it is fully utilized. With virtualization, a physical machine (PM) can host multiple virtual machines (VMs) independently to avoid under-utilization, which enhances energy efficiency at the same time. The approach of combining underutilized servers is called server consolidation. However, non-negligible energy overhead is introduced for the hypervisor software to manage VMs and VM migration. Efficient strategies have been proposed for server consolidation, e.g., a virtual world zone-based strategy for MMORPGs and a resource demand prediction-based strategy for cloud computing. In general, optimization of the mapping between PMs and VMs is formulated as the VM placement problem. Job scheduling determines how efficiently the data center resources are utilized. Joint optimization of VM placement and job scheduling could lead to more significant improvement. Apart from IT equipment, the cooling system also evolves rapidly. For example, the Google AI-empowered cooling system reduced its energy consumption by 40\%.

    \subsection{Artificial Intelligence}
    \label{adv-ai}
    
    \subsubsection{Machine Learning Model}
    
    An efficient model design contributes to reduced computation and considerable energy savings. Certain models inherently have lower power requirements, e.g., spiking  neural networks (SNNs) leverage sparse event-driven spikes and MobileNet adopts depthwise convolution for reducing computation. Compact architecture involves trade-offs on architecture hyperparameters. Increasing complexity results in diminishing marginal returns, and even sometimes hurts performance. Methods, e.g., Neural Architecture Search (NAS), have been developed to automate the design for optimal architectures. 

    Model compression provides an effective way to reduce model sizes and computation with minimal impact on performance for AI inference. For example, pruning  prunes out the non-critical sections (e.g., connections, neurons, and layers) in the neural networks. The energy-aware pruning reduces the AlexNet and GoogLeNet energy consumption by 3.7 and 1.6 times \cite{pruning}. Quantization approximates floating-point numbers by low bit-width numbers for ML model parameters. Model distillation, also called knowledge distillation, trains a small-size "student" model to mimic a large-size "teacher" model. The pruned, quantized, or distilled model is best suited for AI inference at scale. Moreover, dataset distillation distills a large dataset into a synthetic smaller dataset, which provides computation saving for repeated AI training experiments.

    \subsubsection{Computing Paradigm}

    Computation can be performed in different manners, e.g., distributed computing and cloud computing. In distributed computing, AI training or inference workload is fulfilled by multiple worker nodes, which reduces the processing time and mitigates the single point of failure. Software (e.g., DeepSpeed by Microsoft) has been developed to reduce computing power and memory use for AI training.\footnote{\url{https://www.deepspeed.ai/}} Another advantage of distributed computing is that it can leverage diverse local resources. For example, FL can enable workload allocation based on the available intermittent renewable energy to improve overall sustainability. Edge computing pushes the computation towards data sources. As only processed data or extracted information needs to be transmitted, it significantly reduces data volume in the transmission. However, centralized cloud computing possesses the advantage of superior energy efficiency from high-end hardware and centralized facility management. Therefore, the most efficient design could vary from application to application, depending on the factor (e.g., data transmission or computation) that is more dominating in the overall energy consumption. Last but not least, open-sourced pre-trained AI models (e.g., ModelScope by Alibaba) largely encourage the adoption of Pre-train and Fine-tune Paradigm and continual learning, which significantly avoids the energy waste in repeated AI training.\footnote{\url{https://modelscope.cn/}}

    \subsection{Extended Reality}
    \label{adv-xr} 

    \subsubsection{Video Processing}

    Software and algorithmic improvements play a key role in improving video processing efficiency. For instance, data reuse is exploited in the computation of 360$^\circ$ VR pipeline. By memorizing head orientation and establishing a relation between left and right eye projection, some computations can be skipped. Another improvement opportunity lies in video encoding, which is performed before videos are transmitted to compress the video. It effectively reduces the amount of data for storage and transmission. 360$^\circ$ VR videos are currently encoded in three ways: equirectangular projection, cubemap projection, and pyramid projection. Compared to equirectangular projection, cubemap and pyramid projection can save 25\% and 80\% file sizes. 
    
    \subsubsection{Human Visual Perception}

    Characteristics and limitations of human visual perception provide opportunities to improve XR efficiency. Firstly, the fovea, the central area of the human eye, has a substantially higher visual acuity than the rest of the retina. In light of this fact, foveated rendering is developed to render the gazed area of the image at a higher resolution and the peripheral area at a lower resolution. According to \cite{jabbireddy2022foveated}, it can achieve 3 times speedup and 70\% pixel reduction. The same concept can also be incorporated with video compression to reduce the bandwidth and energy required during data transmission. Secondly, the human visual system has limited capability to perceive details of high spatial and temporal frequencies. Rather than rendering all video frames at full resolution, if the resolution of every other fame is reduced, the perceived quality could be still maintained. Multiplex video resolution can also reduce both rendering workload and data transmission. Furthermore, the dark adaptation is another feature of the human eye that increases visual sensitivity in a dark environment. By smoothly decreasing the brightness level of the screen, a significant amount of energy can be saved while keeping the same brightness perception. 

    \subsection{Blockchain}
    \label{adv-blockchain} 

\begin{figure*}[b]
\centering
 \includegraphics[width=15cm, height=8cm,trim={0cm 17cm 0cm 0cm},clip]{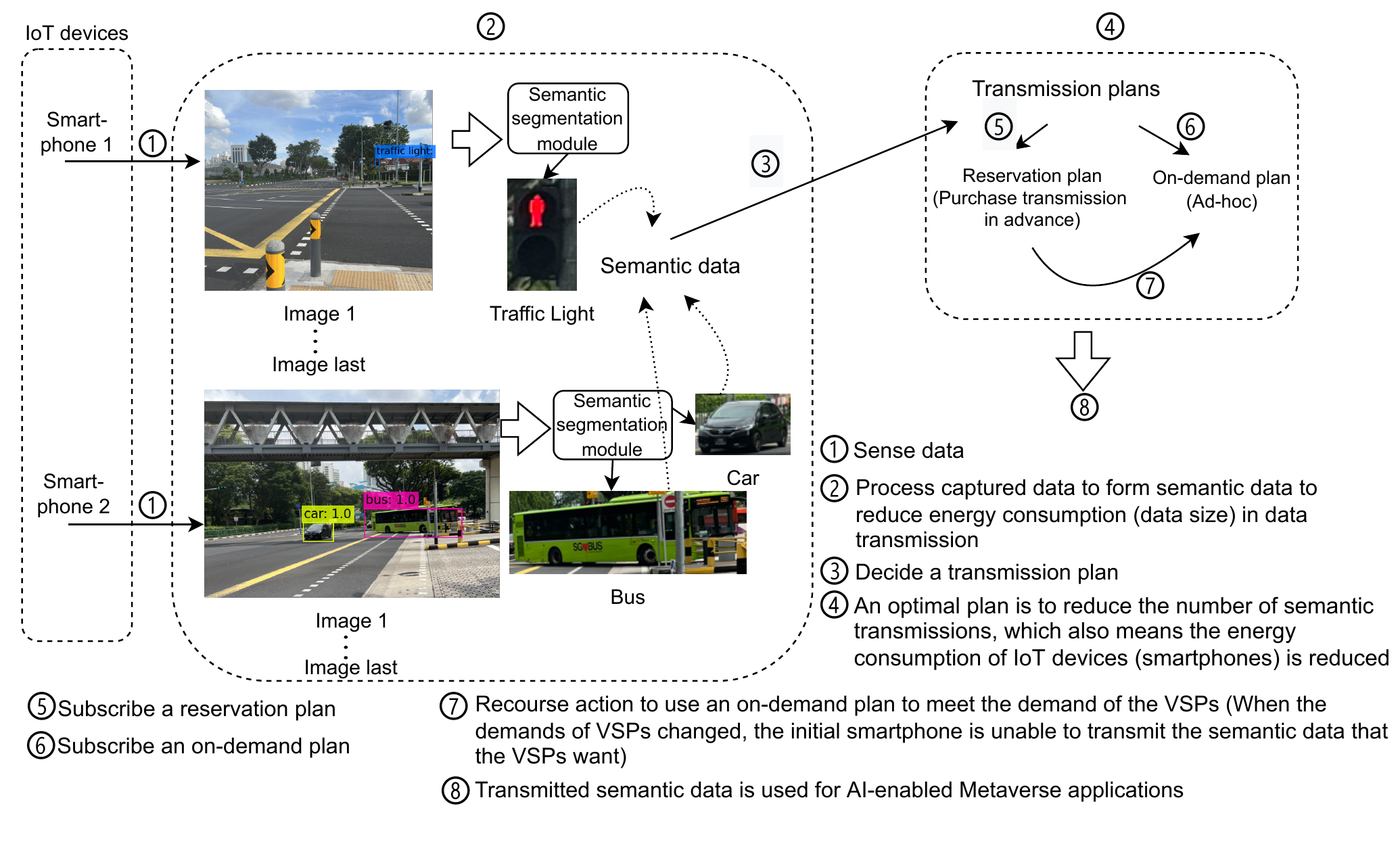}\par
  \caption{System model in the virtual transportation network.}
  \label{fig:systemmodel}
\end{figure*}

    \subsubsection{Consensus Mechanism}
    
    The first type of green advancement lies in alternative energy-efficient consensus mechanisms, e.g., PoS, Delegated PoS (DPoS), and Proof-of-Authority (PoA). PoS replaces miners with validators to validate transactions and add new blocks to the blockchain. Instead of expending assets up-front as energy expenditure, validators stake crypto assets as collateral against dishonest behavior. DPoS, a notable iteration of PoS, allows participants to elect delegates with voting power proportional to their stakes. Only a capped number of delegates are chosen for transaction validation and block production. Thus, it makes consensus more efficient to reach, at the expense of decentrality. While the above mechanisms are open to the public, PoA pre-approves participants with trustworthy identities. Participants stake identities and voluntarily disclose themselves to become validators. The simplicity of consensus-reaching processes and a smaller number of validators make PoA require minimal computational effort. In short, eliminating the mining process (e.g., The Merge by Ethereum) significantly reduces energy consumption by $\sim$99.95\%.\footnote{\url{https://ethereum.org/en/upgrades/merge/}}

    \subsubsection{Operation Redundancy}

    The second type of advancement focuses on reducing operation redundancy, especially for non-PoW blockchains. Reducing the number of nodes to process and the workload in the process can both contribute to redundancy reduction. The sharding technique splits the blockchain network into smaller independent partitions, i.e., shards, and processes each transaction only on nodes in one shard rather than the whole network. Therefore the number of redundant operations can be reduced as well as the energy consumption.
    
    Layer 2 solutions aim to reduce the workload or complexity of the transaction operation by building a second-layer protocol on top of the foundational blockchain. For example, the off-chain payment protocol (e.g., Lightning for Bitcoin and Raiden for Ethereum) creates a direct channel between two participants and enables instantaneous bidirectional transfers, as long as the net sum of transfers does not exceed the deposit. Because the blockchain is involved only when the channel is created or closed, the intensive and redundant computation required for blockchain operations is largely avoided. Zero-knowledge rollups (ZK-rollups) bundle transactions into batches and execute them off-chain. An easily verifiable proof is generated and submitted back to the blockchain, which proves the correctness of changes after executing a batch of transactions. All blockchain participants validate the proof instead of raw transactions, obviously simplifying the workload. Slightly offset by proof generation, ZK-rollup adoption is estimated to reduce 98.5\% of energy demand \cite{papageorgiou2021energy}.

    \textbf{Lessons learned:} Few industries such as telecommunication and data center operators have been intensively working on sustainable development. Green AI and blockchain have taken off in recent years given more widespread adoption. For XR, research on performance improvement, rather than green technologies, still dominates.

\section{Case Study: Stochastic Resource Allocation for Semantic Communication in the Metaverse}
\label{sec4}

\begin{figure}[b]
\centering
    \includegraphics[width=0.8\columnwidth]{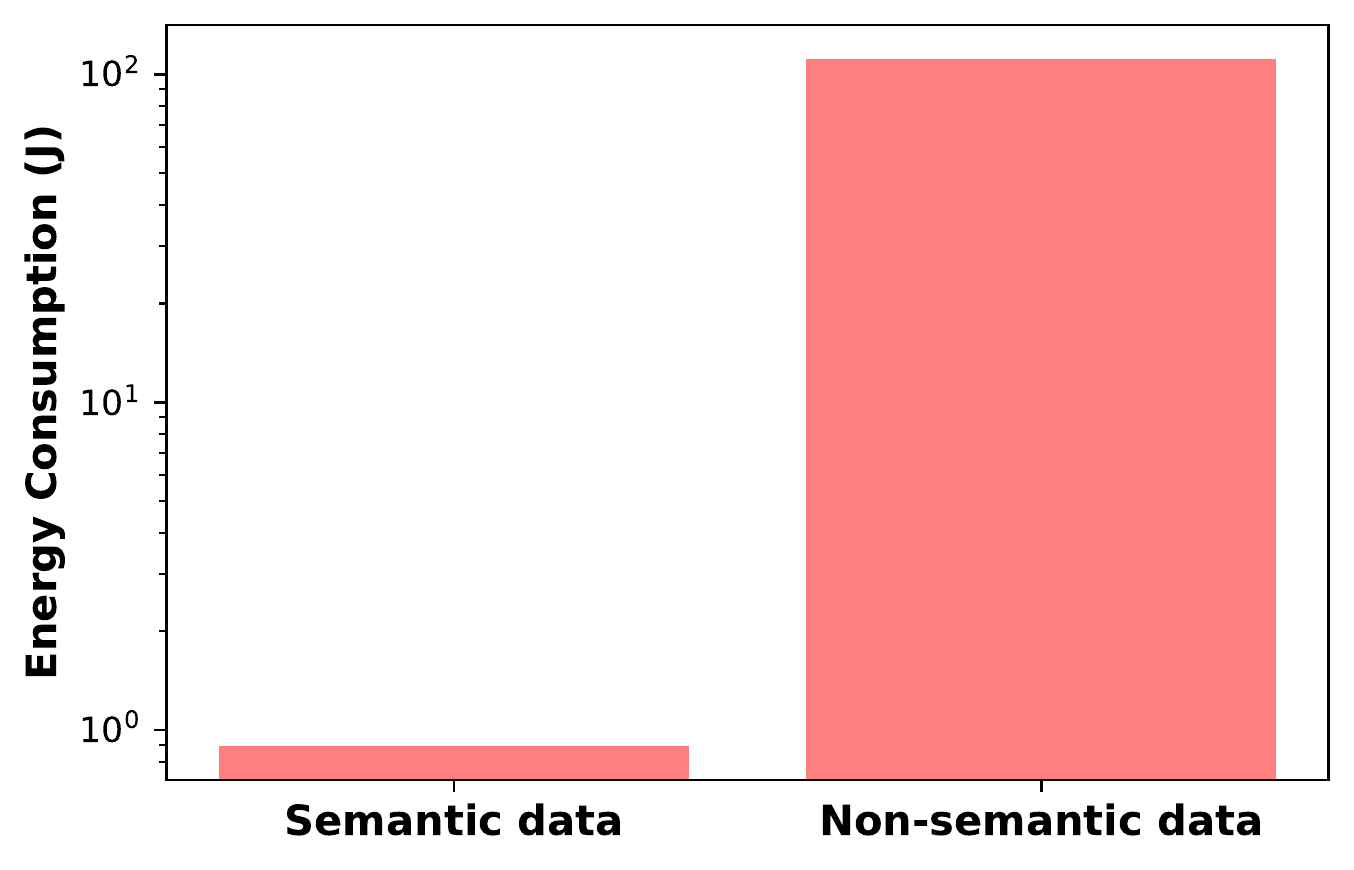}
    \caption{The energy consumption of semantic data and non-semantic data. }
    \label{fig:energy}
\end{figure}

This section presents a case study of the virtual transportation network (VTN) in the Metaverse~\cite{ng2022stochastic}. For example, a transport company can create a VTN to provide on-the-job training for its drivers and to conduct large-scale user studies in a safe but realistic environment. A realistic VTN is developed through the synchronization of sensing data from edge devices in the physical world and DTs of the virtual world. To keep DTs updated, sensing data is regularly transmitted to virtual service providers (VSPs). Fig. \ref{fig:systemmodel} illustrates the system model.

\textbf{Energy-saving semantic communication:} It is energy-consuming to transmit data from edge devices to VSPs. Specifically, the edge device transmits the original image to VSPs, while ignoring the semantic importance of the bit flow. The objective of semantic communication is to be user-oriented, i.e., transmit only the part of semantic data interested by VSPs. For example, an autonomous driving company requires images (vehicles that are driving on the road) to train detection models in the Metaverse. Therefore, instead of transmitting the entire captured images which involve other objects, the semantic extraction module outputs only the segmented objects of interest for transmission. With the help of YOLO, semantic data such as snapshots can be extracted from the captured image. For example, after an image is captured by an edge device, a pre-trained YOLO is used to process and obtain each snapshot of the corresponding object in the image. However, it is unrealistic for the VSPs to search manually to identify which semantic data is within their interest. Therefore, BERT, a powerful pre-trained machine-learning model, is used to convert the objects' labels as well as the interest of the VSPs into vectors according to different contexts. Then, cosine similarity is applied to measure the similarity between the two vectors. With the similarity score, resource allocation can be performed to allocate the resources. We can validate in Fig. \ref{fig:energy} that the energy consumption for semantic data transmission only requires 0.896J as compared to 111J for non-semantic data. Moreover, inferences on pre-trained semantic extraction model such as YOLO requires very little energy \cite{kim2020spiking}. Therefore, with semantic communication, edge devices can reduce power consumption during transmission as well as storage costs.

\begin{figure}[b]
\centering
 \includegraphics[width=0.8\columnwidth]{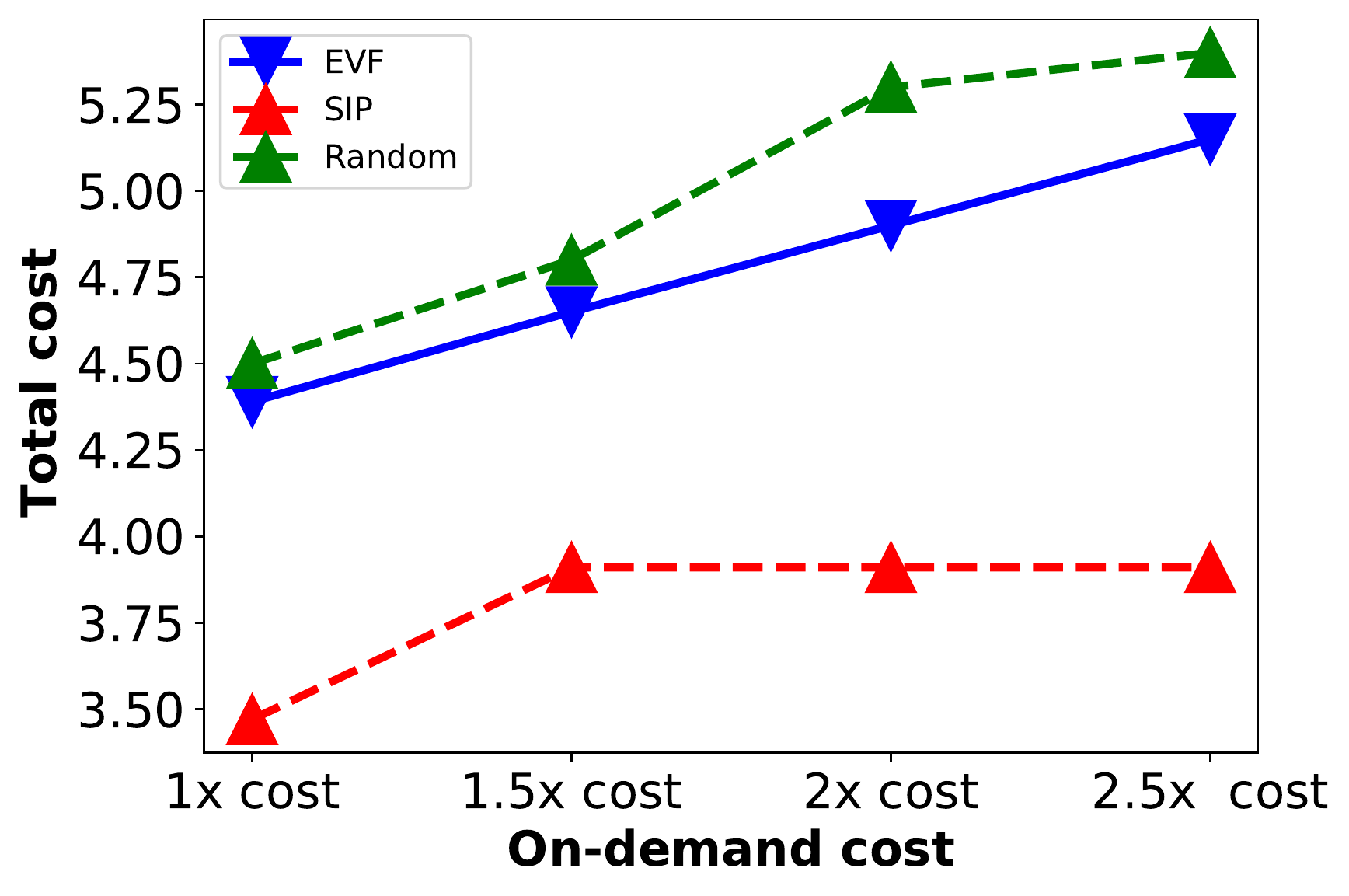}\par
  \caption{SIP comparing with EVF and random schemes.}
  \label{fig:evf}
\end{figure}

\textbf{Stochastic resource allocation for semantic communication:} To maintain DTs for realistic VTNs, VSPs may tap into data marketplaces that aggregate data from edge devices and sell to interested buyers. We assume that the cost of data is proportional to the energy cost incurred by edge devices in the case study. There are two types of subscription plans in data marketplaces, namely "reservation" and "on-demand". The reservation plan allows the VSPs to choose the edge device and purchase the number of data transmissions in advance. We consider two VSPs and three edge devices (smartphones) around Singapore. The reservation cost consists of two components, the membership cost and the transmission cost of one bundle. The VSPs can purchase the number of transmission bundles only when the membership cost is paid. We consider the daily rental cost of smartphones as the membership cost, and there are $n$ semantic data transmissions in one bundle. The on-demand plan is an ad-hoc plan that is only used when needed, so it is more expensive than the reservation plan. However, the demands of VSPs are constantly changing as it is highly dependent on the users' requests. Furthermore, since the reservation plan is subscribed in advance, the demand of the VSPs may not be the same after the subscription. For example, the demand is initially the vehicles on the road. After a while, the detection system may not be able to detect pedestrians very well under extreme weather conditions. Hence, the on-demand plan may be triggered to obtain more data to update the AI model. However, if an incorrect plan is used, e.g., reserve the edge device that transmits semantic data irrelevant to the interest of VSP, energy is wasted from unwanted data transmission, and additional energy is required for re-transmission from other edge devices. 

To account for the demand uncertainties of VSPs, we proposed a two-stage stochastic integer programming method to minimize the operation cost of VSPs as well as transmission energy. The first stage is the reservation stage, which consists of two decision variables, i) membership and ii) bundle variables. The membership variable is a binary variable and it has a corresponding constraint to ensure that the VSP has to pay the membership fee to the edge device owner before the VSP can purchase any bundle from the respective edge device. The bundle variable is a positive variable to indicate the number of bundles to purchase. The second stage is the on-demand stage, with an on-demand decision variable. It indicates the number of semantic data transmissions that VSP is requested on-demand from the edge device after the demand is observed. A demand constraint is used to ensure that the demand is met. Additional semantic data transmission is needed from the on-demand plan (a correction action) whenever the reservation plan has insufficient semantic data transmission to cover the demand. Using VSP historical demand data, our resource allocation scheme achieves a much lower cost than other schemes that do not consider the probability distribution of VSP demand (Fig. \ref{fig:evf}).

\section{Key Challenges and Future Research Directions}
\label{sec5}

\textbf{Challenge 1 -- QoE-Driven Green Metrics:} From Table \ref{tab:table1}, we can observe that most of the sustainability-related metrics are based on conventional QoS factors. In contrast, the recent paradigm shift towards the human-centered design of communication systems will shift the focus from QoS to QoE. Unfortunately, QoE is highly subjective based on user preferences. This is exacerbated by the fact that the Metaverse will feature a diverse range of applications and involve new modalities of user data. As such, before we can design systems to optimize QoE-driven green metrics, it is important to redefine QoE through large-scale user studies.

\textbf{Challenge 2 -- Metaverse Sustainability Index:} Such index is beneficial for Metaverse providers to plan an improvement roadmap, for users to choose green Metaverse products, and for governments to regulate the healthy development of the industry. However, the enormous amount of technologies and stakeholders involved makes it challenging to assess the overall sustainability. Although technologies can be evaluated individually using the metrics in Table \ref{tab:table1}, their contributions to the Metaverse are still vague currently. Practical frameworks and tools will be needed in the future to integrate diverse technological metrics and guide the overall assessment.

\textbf{Challenge 3 -- Trade-Off Balancing System Performance-Energy Consumption:} While research typically focuses on maximization of system performance, e.g., attaining the highest accuracy/fastest inference of AI models or highest FPS in rendering, it is often the case that some applications may not require such extents of high performance. In this case, the system performance can be slightly sacrificed to reduce the carbon footprint or energy per transaction. However, the trade-off between system performance and energy consumption has to be better studied.

\textbf{Challenge 4 -- New Incentive Schemes for Green Technology Adoption:} Incentive mechanisms designed to encourage green technology adoption, e.g., through tax credits or grants, have successfully encouraged the adoption of green office buildings. The next challenge will therefore be how can incentive mechanisms be designed to extend the coverage to the virtual worlds and the enabling technologies of the Metaverse. 

\textbf{Challenge 5 -- System-Level Design:} Diverse technologies are interconnected together in the Metaverse. On one hand, improving one technology might cause additional consumption to other technologies, e.g., compressed sensing and video encoding reduce the data transmission but increase the computation workload. On the other hand, the full-system design could exploit a larger energy-saving potential than focusing on individual subsystems, e.g., synergistically approximating sensing, computing, memory, and communication. Therefore, the system-level design is pivotal for orchestrating technological advancements towards the GMN.

\section{Conclusion}
\label{con}
In this paper, we first present the energy-hungry technologies of the Metaverse. Then, we discuss recent technological advances for the GMN. We subsequently present a case study of the semantic communication-enabled Metaverse development. Finally, we discuss the key challenges and future research directions.

\section*{Acknowledgement}

This research is supported in part by the National Research Foundation (NRF), Singapore and Infocomm Media Development Authority under the Future Communications Research Development Programme (FCP), and DSO National Laboratories under the AI Singapore Programme (AISG Award No: AISG2-RP-2020-019), under Energy Research Test-Bed and Industry Partnership Funding Initiative, part of the Energy Grid (EG) 2.0 programme, and under DesCartes and the Campus for Research Excellence and Technological Enterprise (CREATE) programme. The research is also supported by the SUTD SRG-ISTD-2021-165, the SUTD-ZJU IDEA Grant (SUTD-ZJU (VP) 202102), and the Ministry of Education, Singapore, under its SUTD Kickstarter Initiative (SKI 20210204).

\bibliographystyle{unsrt} % We choose the "plain" reference style
\bibliography{citation} % Entries are in the refs.bib file

\section*{Biography}

SIYUE ZHANG is currently pursuing a Ph.D. degree with Alibaba Group and Alibaba-NTU Joint Research Institute, Nanyang Technological University (NTU), Singapore. His research interests include sustainability, the Metaverse, and artificial intelligence.

WEI YANG BRYAN LIM is currently Wallenberg-NTU Presidential Postdoctoral Fellow. His research interests include edge intelligence and resource allocation.

WEI CHONG NG is currently pursuing a Ph.D. degree with Alibaba Group and Alibaba-NTU Joint Research Institute, Nanyang Technological University, Singapore. His research interests include the Metaverse, stochastic integer programming, and edge computing.

DUSIT NIYATO [IEEE Fellow] is currently a Professor with the School of Computer Science and Engineering and, by courtesy, School of Physical and Mathematical Sciences, Nanyang Technological University, Singapore. His research interests are sustainability, edge intelligence, decentralized machine learning, and incentive mechanism design.

XUEMIN (SHERMAN) SHEN [IEEE Fellow] is currently a university professor with the Department of Electrical and Computer Engineering, University of Waterloo. His research focuses on network resource management, wireless network security, social networks, 5G and beyond, and vehicular ad hoc networks.

CHUNYAN MIAO [IEEE Fellow] is currently a professor in the School of Computer Science and Engineering, Nanyang Technological University (NTU), and the director of the Joint NTU-UBC Research Centre of Excellence in Active Living for the Elderly (LILY).

\end{document}